\documentclass[letterpaper, 10 pt, conference]{ieeeconf}  
\usepackage{graphicx}    
\usepackage{comment}  
\usepackage[utf8]{inputenc}
\usepackage{amsmath}
\usepackage{amssymb}
\newcommand{\norm}[1]{\lVert#1\rVert}
\usepackage[dvipsnames]{xcolor}
\usepackage{multirow}
\usepackage{accents}
\usepackage{subcaption}

\usepackage{amsmath}

\IEEEoverridecommandlockouts  

\title{\LARGE \bf
	Neural Network Adaptive Control with Long Short-Term Memory}
\author{Emirhan Inanc, Yigit Gurses, Abdullah Habboush, Yildiray Yildiz and Anuradha M. Annaswamy
	\thanks{E. Inanc, A. Habboush and Y. Yildiz are with the Department of Mechanical Engineering, and Y. Gurses is with the Department of Computer Engineering, Bilkent University, Cankaya, Ankara 06800, Turkey. A.M. Annaswamy is with the Department of Mechanical Engineering, Massachusetts Institute of Technology, Cambridge, MA, 02139. (e-mails: emirhan.inanc@bilkent.edu.tr, yigit.gurses@ug.bilkent.edu.tr,  a.habboush@bilkent.edu.tr, yyildiz@bilkent.edu.tr, aanna@mit.edu).}
}

\begin{document}
	
	\maketitle
	\thispagestyle{empty}
	\pagestyle{empty}

	\begin{abstract}
		In this study, we propose a novel adaptive control architecture, which provides dramatically better transient response performance compared to conventional adaptive control methods. What makes this architecture unique is the synergistic employment of a traditional, Adaptive Neural Network (ANN) controller and a Long Short-Term Memory (LSTM) network. LSTM structures, unlike the standard feed-forward neural networks, can take advantage of the dependencies in an input sequence, which can contain critical information that can help predict uncertainty. Through a novel training method we introduced, the LSTM network learns to compensate for the deficiencies of the ANN controller during sudden changes in plant dynamics. This substantially improves the transient response of the system and allows the controller to quickly react to unexpected events. Through careful simulation studies, we demonstrate that this architecture can improve the estimation accuracy on a diverse set of uncertainties for an indefinite time span. We also provide an analysis of the contributions of the ANN controller and LSTM network to the control input, identifying their individual roles in compensating low and high-frequency error dynamics. This analysis provides insight into why and how the LSTM augmentation improves the system's transient response. The stability of the overall system is also shown via a rigorous Lyapunov analysis.
		
	\end{abstract}
	
	\section{Introduction}
	\label{sec:introduction}
		
	Although Neural Networks (NN) is a fairly old concept \cite{macukow2016neural}, cheap and fast parallel computing unlocked their potential and lead to their current predominance in artificial intelligence and machine learning \cite{aggarwal2018neural}. Computer vision and natural language processing are examples of fields that benefited greatly from these developments \cite{russakovsky2015imagenet}, \cite{otter2020survey}. Deep learning research produced derivatives of recurrent neural networks (RNN) such as long short-term memory (LSTM) \cite{van2020review} and gated recurrent unit \cite{cho2014properties}, which are powerful structures for processing sequential data \cite{lipton2015critical}. Reinforcement learning (RL) is another concept in machine learning that made advancements through applications of deep learning \cite{arulkumaran2017brief}. In RL, there exists a state that is updated with respect to an action selected by an agent that makes its choices based on observations to maximize its cumulative reward. A similarity can be drawn between control systems and RL where states, control inputs and feedback connections are parallel to states, actions, and observations, respectively. There are studies that take advantage of this fact to design RL-based controllers \cite{lewis2012reinforcement}, \cite{bucsoniu2018reinforcement}.
	
	The potential shown by earlier applications provides an incentive to utilize NN in adaptive control. The literature on this topic is extensive and well established \cite{calise2001adaptive,chen2001nonlinear,kwan2000robust,lewis1996multilayer,annaswamy2021historical}. Using an online-tuned feed-forward NN controller with adaptive update laws is proven to be stable and can successfully compensate the uncertainties \cite{lewis1996multilayer}. However, if there are sudden changes in the uncertainty, the transient response can be oscillatory, which results in poor performance. This problem is addressed in \cite{muthirayan2022memory} with the addition of external memory similar to a neural turing machine's (NTM) \cite{graves2014neural}. However, unlike the common practice in NTMs, a feed-forward network, instead of an RNN, is used in [17]. Feed-forward networks that have access to only the current state cannot take full advantage of the dependencies in a sequence. Instead, using a recurrent structure with internal memory can increase the capability of sequence estimation and thus improve performance \cite{somu2020hybrid}.
	
	To address the above mentioned issues, we propose a novel control architecture that consists of an Adaptive Neural Network (ANN) controller and an LSTM. The purpose of the LSTM is to take advantage of the long and short-term dependencies in the input sequence to improve the transient response of the controller. Specifically, we train the LSTM in such a way that it learns to compensate for the inadequacies of the ANN controller in response to sudden and unexpected uncertainty variations. Offline training in closed-loop systems is challenging since the trained element affects the system dynamics. This closed loop structure similarly exists in RL applications and is addressed by previous studies \cite{thuruthel2018model,kalashnikov2018scalable}. Inspired by these methods, we train the LSTM in a closed-loop setting to predict and compensate for the undesired transient error dynamics. We demonstrate via simulations that thanks to its predictive nature, LSTM offsets high-frequency errors, and thus complements the ANN controller, where the latter helps with handling the low-frequency dynamics.
	
	To summarize, the contribution of this paper is a novel adaptive control framework that provides enhanced transient performance compared to conventional approaches. This is achieved by making the traditional ANN controller work in collaboration with an LSTM network that is trained in the closed-loop system.
	
	In Section \ref{problem formulation}, we describe the formulation of the ANN controller. In Section \ref{Controller Design}, the proposed LSTM augmentation and the training method are explained, together with a stability analysis. Simulation results are given in Section \ref{Simulations} and a summary is given in Section \ref{Summary}.
	
	
	\section{Problem Formulation}\label{problem formulation}
	Consider the following plant dynamics
	\begin{subequations} \label{plant dynamics}
		\begin{align}
			\dot{x}_p(t)=&A_px_p(t)+B_p(u(t)+f(x_p(t))), \\
			y_p(t)=&C_p^Tx_p(t),
		\end{align}
	\end{subequations}
	where $x_p(t)\in \mathbb{R}^{n_p}$ is the accessible state vector, $u(t)\in \mathbb{R}^m$ is the plant control input, $A_p \in \mathbb{R}^{n_p\times n_p}$ is a known system matrix, $f(x_p(t)):\mathbb{R}^{n_p}\xrightarrow{}\mathbb{R}^{m}$ is a state-dependent, possibly nonlinear, matched uncertainty, $B_p \in \mathbb{R}^{n_p\times m}$ is a known control input matrix, and $C_p\in \mathbb{R}^{n_p\times s}$ is a known output matrix, and $y_p(t)\in \mathbb{R}^s$ is the plant output. Furthermore, it is assumed that the pair $(A_p, B_p)$ is controllable.
	
	\textit{Assumption 1:} The unknown matched uncertainty $f(x_p(t)):\mathbb{R}^{n_p}\xrightarrow{}\mathbb{R}^{m}$ is continuous on a known compact set $S_p\triangleq\{x_p(t):\norm{x_p(t)}\leq b_x\}\subset\mathbb{R}^{n_p}$, where $b_x$ is a known positive constant.
	
	\textit{Remark 1:} The proposed method also applies for an unknown state matrix $A_p$: Suppose that the plant dynamics are given by 
	\begin{equation} \label{plant dynamics with unknown Au}
		\dot{x}_p(t)=A_ux_p(t)+B_p(u(t)+f_u(x_p(t))),
	\end{equation}
	where $A_u\in\mathbb{R}^{n_p\times n_p}$ is an unknown state matrix and $f_u(x_p(t))$ is a state-dependent, possibly nonlinear, matched uncertainty. The dynamics (\ref{plant dynamics with unknown Au}) can be written in the form of (\ref{plant dynamics}) and for a known matrix $A_p$, by defining $f(x_p(t))\triangleq K_ux_p(t)+f_u(x_p(t))$, given that the matching condition $A_p=A_u-B_pK_u$ is satisfied for some feedback gain $K_u\in\mathbb{R}^{m\times n_p}$. This allows us to proceed assuming $A_p$ is known, without loss of generality.

	To achieve command following of a bounded reference input $r(t)\in\mathbb{R}^s$, an error-integral state $x_e(t)\in\mathbb{R}^s$ is defined as
	\begin{equation} \label{integral action}
		\dot{x}_e(t)=r(t)-y_p(t),
	\end{equation}
	which, when augmented with (\ref{plant dynamics}), yields the augmented dynamics
	\begin{subequations}\label{augmented_dynamics}
		\begin{align}
			\dot{x}(t)&=Ax(t)+B_mr(t)+B(u(t)+f(x_p(t))), \label{augmented_dynamics_a} \\
			\dot{y}(t)&=C^Tx(t), \label{augmented_dynamics_b}
		\end{align}
	\end{subequations}
where $x(t)\triangleq[x_p(t)^T, x_e(t)^T]^T\in\mathbb{R}^{n}$ is the augmented state, whose dimension is $n \triangleq n_p+s$, and the system matrices $A$, $B_m$, $B$ and $C$ are
	\begin{subequations} \label{withoutp}
		\begin{align}
			A \triangleq& \begin{bmatrix} A_p & 0_{n_p \times s} \\ -C_p^T & 0_{s \times s} \end{bmatrix} \in \mathbb{R}^{n\times n},\label{A} \\	
			B_m \triangleq& \begin{bmatrix}  0_{n_p \times s} \\ -I_{s\times s} \\ \end{bmatrix} \in \mathbb{R}^{n \times s}, \label{B}\\		
			B \triangleq& \begin{bmatrix}  B_p \\ 0_{s \times m} \\ \end{bmatrix} \in \mathbb{R}^{n \times m}, \label{B}\\
			C \triangleq& \begin{bmatrix} -C_p^T & 0_{s \times s} \end{bmatrix}^T \in \mathbb{R}^{n \times s}. \label{C}
		\end{align}
	\end{subequations}
	A baseline state feedback controller is designed as
	\begin{equation}\label{baseline}
		u_{bl}(t)=-Kx(t),
	\end{equation}
	where $K\in\mathbb{R}^{m\times n}$ is selected such that $A_m\triangleq A-BK$ is Hurwitz. Then, a reference model is defined as
	\begin{equation} \label{reference model}
		\begin{aligned}
			&\dot{x}_m(t)=A_mx_m(t)+B_mr(t),
		\end{aligned}
	\end{equation}
	where $x_m(t)\in \mathbb{R}^{n}$ is the state vector of the reference model.
	
	Although the matched uncertainty $f(x_p(t))$ is of unknown structure, Assumption 1 implies that it can be approximated by a multi-layer neural network in the following form
	\begin{equation} \label{approximation1}
		f(x_p(t)) = W^T\bar{\sigma}(V^T\bar{x_p}(t))
		+\varepsilon{(x_p(t))},
	\end{equation}
	such that
	\begin{equation} \label{eps}
		\norm{\varepsilon{(x_p(t))}}\leq\varepsilon_N, \quad \quad \forall x \in S,
	\end{equation}
	respectively, where the NN reconstruction error vector $\varepsilon{(x_p(t))}$ is unknown, but bounded by a known constant $\varepsilon_N>0$ in the domain of interest $S_p\subset\mathbb{R}^{n_p}$. Such a bound depends on the number of hidden neurons $n_h$ and the weight matrices of the neural network $W$ and $V$. The following assumption is necessary and is standard in the literature.
	
	\textit{Assumption 2:} The unknown ideal weights $W$ and $V$ are bounded by known positive constants such that $\norm{V}_F\leq V_M$ and $\norm{W}_F\leq W_M$.
	
	The input to the NN and the output of the hidden layer is given by
	\begin{subequations}
		\begin{align}
			\bar{x}_p(t) &\triangleq 
			\begin{bmatrix}
			x_p(t)^T & 1 
			\end{bmatrix}^T\in\mathbb{R}^{n_p+1},\\
			\bar{\sigma}(V^T\bar{x}_p(t))&\triangleq
			\begin{bmatrix}
				\sigma(V^T\bar{x}_p(t)) & 1
			\end{bmatrix}^T\in\mathbb{R}^{n_h+1},
		\end{align}
	\end{subequations}
	 where the nonlinear activation function $\sigma(.):\mathbb{R}^{n_h}\to\mathbb{R}^{n_h}$ can be either sigmoid or tanh. Note that $x_p(t)$ and $\sigma(V^T\bar{x}_p(t))$ are augmented with unity elements to account for hidden and outer layers biases, respectively. For the convenience of notation, we drop the overbar notation and write (\ref{approximation1}) as
	\begin{equation} \label{approximation}
		f(x_p(t)) = W^T\sigma(V^Tx_p(t))+\varepsilon{(x_p(t))},
	\end{equation}
	where the dimensions of the weight matrices are loosely defined as $W\in\mathbb{R}^{n_h\times m}$ and $V\in\mathbb{R}^{n_p\times n_h}$.
	
	By defining the state tracking error as
	\begin{equation} \label{tracking_error}
		e(t)\triangleq x(t)-x_m(t),
	\end{equation}
	the control objective is to make $e(t)$ converge to zero while keeping all system signals bounded, with minimal transients. To achieve this, we use the control input
	\begin{equation} \label{control input}
		u(t) = u_{bl}(t)+u_{ad}(t)+u_{lstm}(t)+v(t),
	\end{equation}
	where $u_{bl}(t)$ is the baseline controller (\ref{baseline}) that achieves the reference model dynamics (\ref{reference model}) in the absence of uncertainty, $u_{ad}(t)$ is an adaptive neural network (ANN) controller designed to compensate for the matched uncertainty $f(x_p(t))$, and $u_{lstm}$ is the output of a long short-term memory (LSTM) network. Furthermore, $v(t)$ is a robustifying term, which is defined subsequently. 
	
	In this paper, we focus our attention on enhancing the transient response of the closed-loop system. Particularly, we propose the synergistic employment of a traditional ANN controller and an LSTM network to obtain superior performance compared to conventional approaches. In what follows, we detail each control term in (\ref{control input}).

	\subsection{Adaptive Neural Network Controller} \label{ANN}
	The adaptive neural network (ANN) controller is designed to compensate for the matched nonlinear uncertainty $f(x_p(t))$, and is expressed as,
	\begin{equation} \label{u_ad}
		u_{ad}(t) = -\hat{f}(x_p(t)),
	\end{equation}
	where $\hat{f}(x_p(t))$ represents an estimate of $f(x_p(t))$. Using (\ref{approximation}), we define the estimate $\hat{f}(x_p(t))$ as
	\begin{equation} \label{fhat}
		\hat{f}(x_p(t)) = \hat{W}^T(t)\sigma(\hat{V}^T(t)x_p(t)),
	\end{equation}
	where $\hat{V}(t) \in \mathbb{R}^{n_h\times n}$ is the input-to-hidden-layer weight matrix and $\hat{W}(t) \in \mathbb{R}^{m\times n_h}$ is the hidden-to-output-layer weight matrix. These weights serve as estimates for the unknown ideal weights $V$ and $W$, respectively. Through proper use of Taylor series-based arguments \cite{lewis1996multilayer}, the weights are updated using the adaptive laws 
	\begin{subequations} \label{NN weight updates}
		\begin{align}
			\dot{\hat{W}} &= F(\hat{\sigma}-\hat{\sigma}'\hat{V}^Tx_p)e^TPB- F\kappa\norm{e}\hat{W}, \label{NN weight updates_W}
			\\
			\dot{\hat{V}} &= Gx_pe^TPB\hat{W}^T\hat{\sigma}'-G\kappa\norm{e}\hat{V}, \label{NN weight updates_V}
		\end{align}
	\end{subequations}
	where $F\in\mathbb{R}^{n_h\times n_h}$ and $G\in\mathbb{R}^{n_p\times n_p}$ are symmetric positive definite matrices, serving as learning rates, $\kappa>0$ is a scalar gain, and $\hat{\sigma}$ and $\hat{\sigma}'$ are defined as
	
	\begin{equation} \label{sigma}
		\begin{aligned}
			&\hat{\sigma}\triangleq\sigma(\hat{V}^T(t)x_p(t)),
			\\
			&\hat{\sigma}'\triangleq \frac{d\sigma(z)}{dz}\bigg|_{z={\hat{V}}^T(t)x_p(t)}.
		\end{aligned}
	\end{equation}
	Further, $P \in \mathbb{R}^{n\times n}$ is the symmetric positive definite solution of the Lyapunov equation
	\begin{equation}\label{lyap eqn}
		A_{m}^TP+PA_{m} = -Q, 
	\end{equation}
	for some symmetric positive definite matrix $Q \in \mathbb{R}^{n\times n}$.
	
	\subsection{Robustifying Term}
	
	The usage of the baseline controller, ANN controller, and the robustifying term is sufficient to solve the adaptive control problem, as demonstrated in \cite{lewis1996multilayer} for nonlinear robot arm dynamics. However, for a generic multi-input multi-output (MIMO) dynamical system with matched uncertainties (\ref{plant dynamics}), a difficulty arises from the control input matrix $B$. A subsidiary contribution of this paper is a modified robustifying term $v(t)$, which achieves the same stability results in \cite{lewis1996multilayer}, but for generic MIMO linear systems.
	
	The robustifying term, $v(t)$ (see (\ref{control input})) is used to provide robustness against disturbances that arise from the high-order Taylor series terms \cite{lewis1996multilayer}. For the considered MIMO linear plant dynamics, we propose a modified robustifying term which is defined as
	\begin{equation} \label{robustifying_term}
		v(t) = 
		\begin{cases} 
			0, & \text{if } \norm{B^TPe}=0\\
			-\frac{B^TPe}{\norm{B^TPe}}k_z\norm{e}(\norm{\hat{Z}}_F+Z_M), & \text{otherwise}
		\end{cases},
	\end{equation}
	with
	\begin{equation} \label{gain}
		\begin{aligned}
			k_z>C_2,
		\end{aligned}
	\end{equation}
	where $Z_M\triangleq \sqrt{W_M^2+V_M^2}$, and matrices $B$ and $P$ are defined in the previous subsection, whereas $e$ is given in (\ref{tracking_error}). Furthermore, $C_2\triangleq 2C_{\sigma'}Z_M$, where $C_{\sigma'}$ is a known upper bound for the absolute value of the sigmoid or tanh activation function derivatives, i.e. $||\sigma'(.)|| \leq C_{\sigma'}$, where $\sigma'(z)=d\sigma(z)/dz$.
	
	In order not to diverge from the main contribution of the paper, we defer the derivation of the ANN controller and the proposed modified robustifying term to the Appendix. The following section introduces the main focus of this paper.
	
	\section{LSTM Network Design}\label{Controller Design}
\begin{figure}[t]
		\centering
		\includegraphics[width=\linewidth]{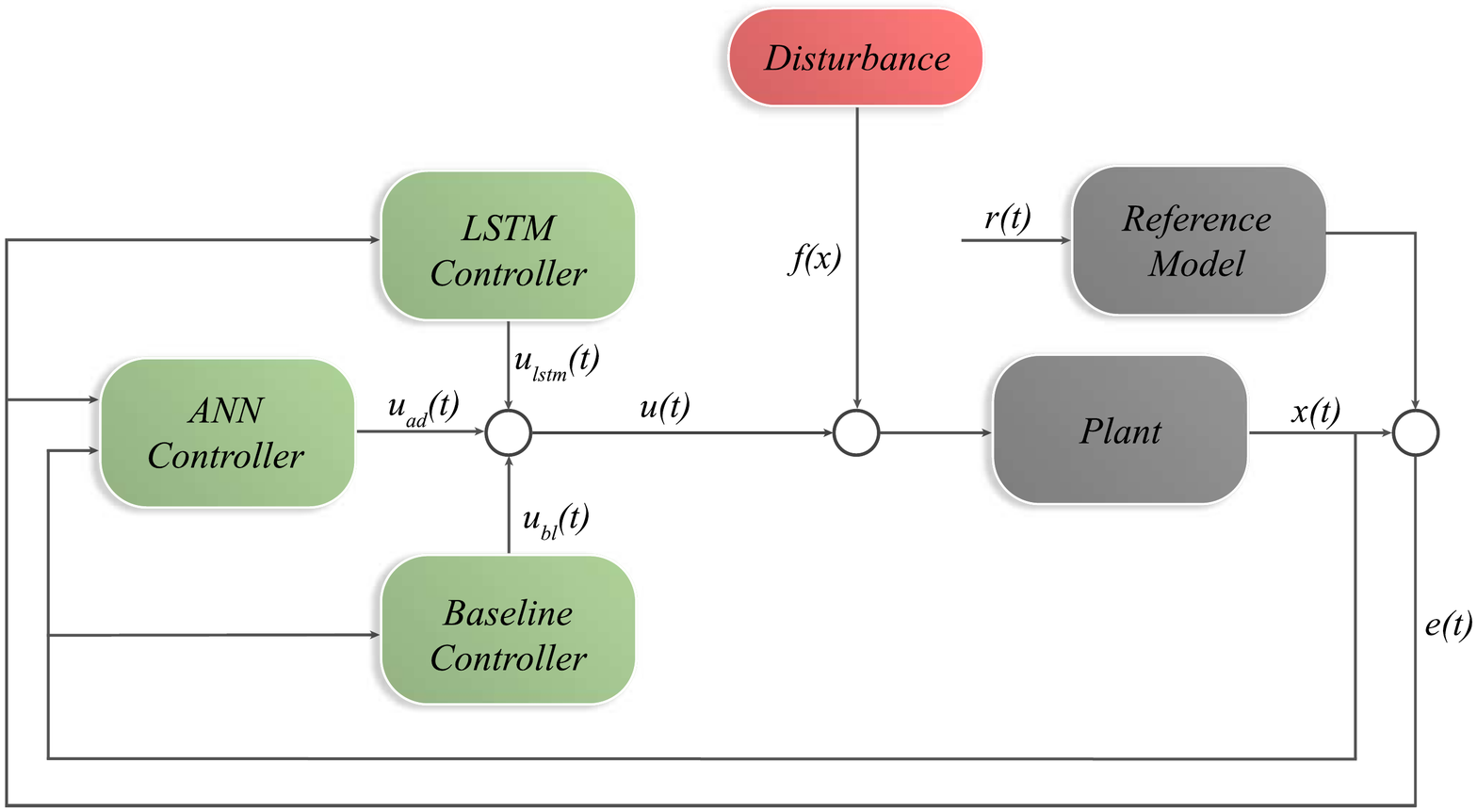}
		\caption{Block diagram of the proposed control architecture.}
		\label{block_digram}
	\end{figure}

	We propose an LSTM network that works in coordination with the ANN controller to compensate for the uncertainties in the system, with better transients compared to conventional approaches. Combined usage of LSTM and ANN is not arbitrary. Indeed, in this section, we precisely define the separate roles of LSTM and ANN in the closed-loop system. The overall control architecture is given in Fig. \ref{block_digram}. By taking the state tracking error $e(t)=x(t)-x_m(t)$ as the input, the LSTM network computes the control input $u_{lstm}(t)$, which is fed to the plant to compensate for the ANN controller deficiencies. The motivation behind the LSTM network is to improve the transient response of the system and achieve faster convergence by taking advantage of the sequence prediction capabilities of the LSTM architecture. 
	
	\textit{Remark 2:} LSTM network takes a sequence as an input and computes a sequence as the output. Hence, the sampling time of the network arises as a design parameter. The particular selection of the sampling time of the network is application dependent. A proper selection would match the sampling time at which the ANN is implemented, or it can be made larger to reduce the computational load. Nevertheless, the exact sampling time of the network is irrelevant in the proposed architecture since, as shown subsequently, the LSTM network can be trained offline to perform under a given sampling time.
	
	\textit{Remark 3:} The LSTM input sequence $e^n$ is generated from the continuous-time signal $e(t)$ as
		\begin{equation}\label{A2D}
			e^n=e(nT),\:\:\:\: n=0,1,2,\dots
		\end{equation}
	where $T>0$ is the sampling time of the LSTM. Furthermore, the output sequence of the LSTM $u_{lstm}^n$ is converted to the continuous-time signal $u_{lstm}(t)$ which accounts for the control input of the LSTM, where 
	\begin{equation}\label{D2A}
		u_{lstm}(t)=u_{lstm}^n, \:\:\:\: \text{for}\:\:  nT\leq t< (n+1)T,\:\:\:\: n=0,1,2,\dots
	\end{equation}
	
\begin{figure}[thpb]
		\centering
		\includegraphics[width=\linewidth]{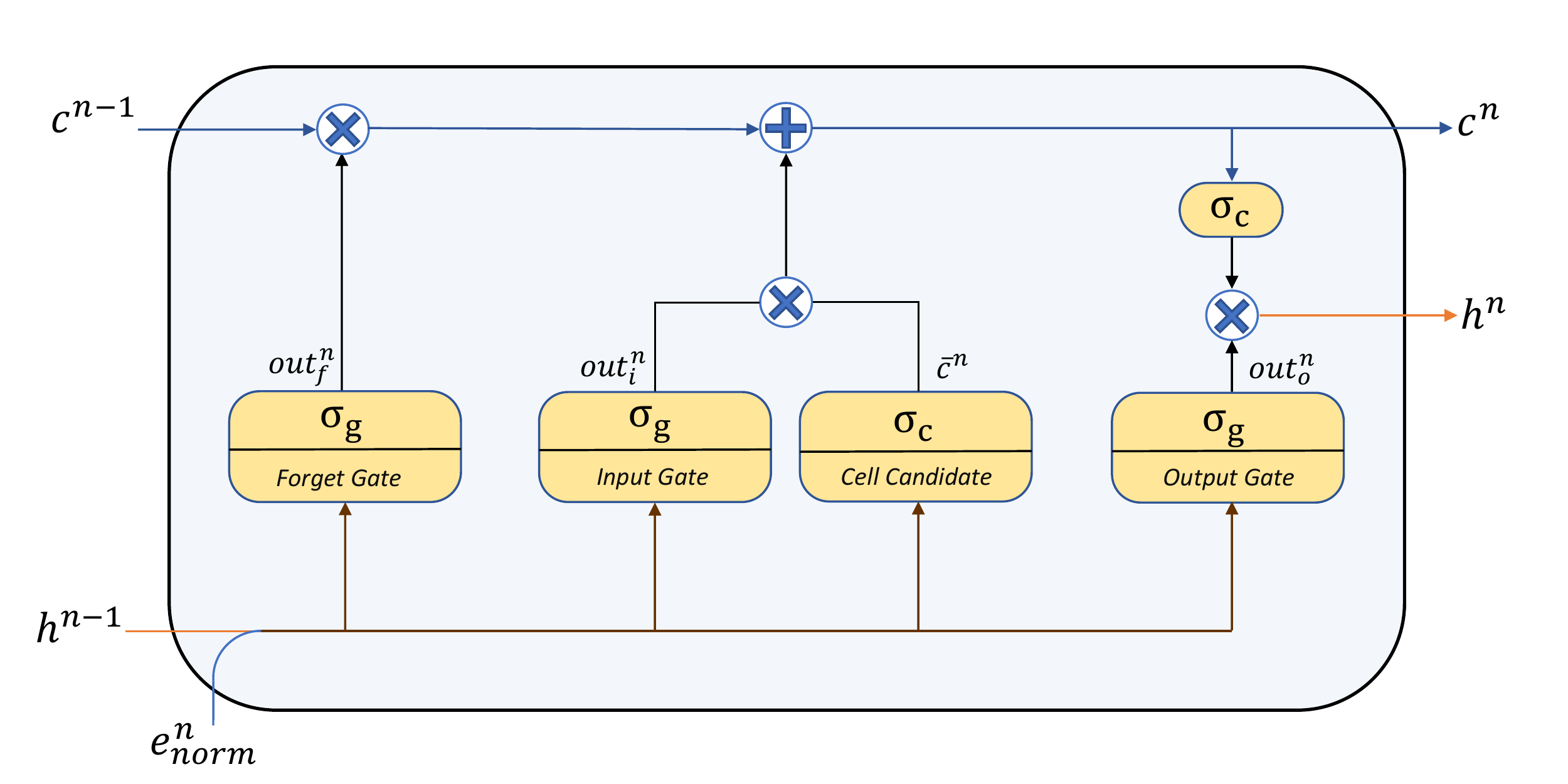}
		\caption{Detailed sketch of an LSTM cell}
		\label{lstm}
	\end{figure}

	LSTM is a Recurrent Neural Network (RNN), which, unlike a conventional neural network structure, can learn the long-term dependencies of a sequence. This is accomplished with an additional hidden state, a cell state, and gates that update the states in a systematic way. One cell of LSTM can be seen in Fig. \ref{lstm}.
	
	In Fig. \ref{lstm}, $e_{norm}^n$ is the normalized version of the input sequence $e^n$, $h^n$ is the hidden state, and $c^n$ is the cell state. $R, W$, and $b$ are the recurrent weight matrices, input weight matrices, and bias vectors, respectively. $\sigma_{g}$ denotes the gate activation function (sigmoid) and $\sigma_c$ denotes the state activation function ($tanh$). Subscripts $f, g, i$, and $o$ denote forget gate, cell candidate, input gate, and output gate, respectively. $out_f^n, \bar{c}^n, out_i^n$, and $out_o^n$ denote the output of the gate operations. The superscript ${n}$ denotes the current time step and ${n-1}$ is the previous time step.
	
	a) Forget Gate: The forget gate decides the relevant information that the cell state should take from the hidden state and the input. This is achieved with a sigmoid activation function.
	
	\begin{equation} \label{forget gate}
		\begin{aligned}
			&out_f^n=\sigma_g(W_fe_{norm}^{n}+R_fh^{n-1}+b_f).
		\end{aligned}
	\end{equation}
	
	b) Input Gate: The new information that is going to be added to the cell state is determined by the input gate and cell candidate. First, the hyperbolic tangent $tanh$ activation function is used to determine which information is going to be added to the cell state. Then, a sigmoid function is used to determine how much of this information is going to be added. This process is given as
	
	\begin{equation} \label{cell candidate}
		\begin{aligned}
			&\bar{c}^n=\sigma_c(W_ge_{norm}^{n}+R_gh^{n-1}+b_g),
		\end{aligned}
	\end{equation}
	
	\begin{equation} \label{input gate}
		\begin{aligned}
			&out_i^n=\sigma_g(W_ie_{norm}^{n}+R_ih^{n-1}+b_i).
		\end{aligned}
	\end{equation}
	
	c) Output Gate: The output gate decides how much of the cell state is going to be a part of the hidden state, using a sigmoid function as
	\begin{equation} \label{output gate}
		\begin{aligned}
			&out_o^n=\sigma_g(W_oe_{norm}^{n}+R_oh^{n-1}+b_o).
		\end{aligned}
	\end{equation}
	The cell state $c$ is first erased using the forget gate $out_f$, then new information is added through the input gate $out_i$ and cell candidate $\bar{c}$. This is achieved with the calculation
	\begin{equation} \label{cell state}
		\begin{aligned}
			&c^n=out_f^n\odot c^{n-1}+out_i^n \odot \bar{c}^n.
		\end{aligned}
	\end{equation}
	The gates mentioned above are used to determine the hidden state $h$ and the cell state $c$ for the next time step. The hidden state is updated using the output gate $out_o$ and cell state $c$ that is mapped between -1 and 1 using a $tanh$ function as
	\begin{equation} \label{hidden state}
		\begin{aligned}
			&h^n=out_o^n \odot \sigma_{c}(c^n).
		\end{aligned}
	\end{equation}
	A fully connected layer is utilized to obtain an $m-$dimensional output to match the dimension of the control input as
	
	\begin{equation} \label{u_lstm}
		\begin{aligned}
			&u_{lstm}^n=W_{fc}h^n+b_{fc},
		\end{aligned}
	\end{equation}
	where $W_{fc}$ and $b_{fc}$ are the weight matrix and bias vector of the fully connected layer, respectively. Then, the control input $u_{lstm}(t)$ is constructed from the LSTM output as shown in (\ref{D2A}).
	
	The ability to learn the dependencies within sequences helps the LSTM to predict the evolution of an unknown function. It is noted that in a typical ANN implementation, the speed of the response can be increased by increasing the adaptation rates, which may cause undesired oscillations. With LSTM augmentation, faster convergence is achieved without the need for large learning rates. The training and implementation details of LSTM are explained in the following subsections.

	\subsection{Training Method}
	
	Since the LSTM network operates in a closed loop system, the system states and the ``truth'', which is the variable it is trying to estimate, are affected by the LSTM output (see Fig. \ref{block_digram}). As the training progresses, the training data itself evolves with time, which allows the data to be used only once to train the network. 

	Our main philosophy to use LSTM is to compensate for the inadequacies of the conventional ANN controller: We know that ANN controllers are successful in compensating for the uncertainties asymptotically. However, adjusting their transient characteristics is not a trivial task. Instead of increasing the learning rates, LSTM provides better transients by actually predicting the type of transients and compensating accordingly. To achieve this, the goal for the LSTM is set to predict the estimation error of the ANN controller and compensate for this error. The estimation error of the ANN, or the deviation of the ANN controller input (\ref{u_ad}), $\hat{f}(x_p(t))$, from the actual uncertainty, $f(x_p(t))$, is given as
	\begin{equation} 
		\begin{aligned}
			&\tilde{f}(x_p(t))\triangleq{f}(x_p(t))-\hat{f}(x_p(t)).
		\end{aligned}
	\end{equation}
	The ``truth'' for the LSTM, or the desired signal for the LSTM to produce, is selected as
	\begin{equation} \label{lstmtraining}
		\begin{aligned}
			&y(t) = -\tilde{f}(x_p(t)).
		\end{aligned}
	\end{equation}
	Therefore, LSTM provides a signal to the system that is an estimate of this truth, expressed as   
	\begin{equation} \label{truth}
		\begin{aligned}
			u_{lstm}(t) = \hat{y(t)}.
		\end{aligned}
	\end{equation}

	During training, an uncertainty $f_{train}(x_p(t))$ is introduced to the closed loop system, and LSTM is expected to learn $\tilde{f}_{train}(x_p(t))=f_{train}(x_p(t))-\hat{f}(x_p(t))$, where $-\hat{f}(x_p(t))$ is the control signal produced by ANN. Apart from $\tilde{f}_{train}(x_p(t))$, LSTM uses the state-tracking error $e(t)$, defined in (\ref{tracking_error}), as its input (see Fig. \ref{block_digram}). Every iteration of training affects the sequence that LSTM is trained on. This dynamic nature of training helps LSTM generalize to a set of functions that are not used in the training set. This is further emphasized in the Simulations Section. 
		
\begin{figure}[thpb]
		\centering
		\includegraphics[width=\linewidth]{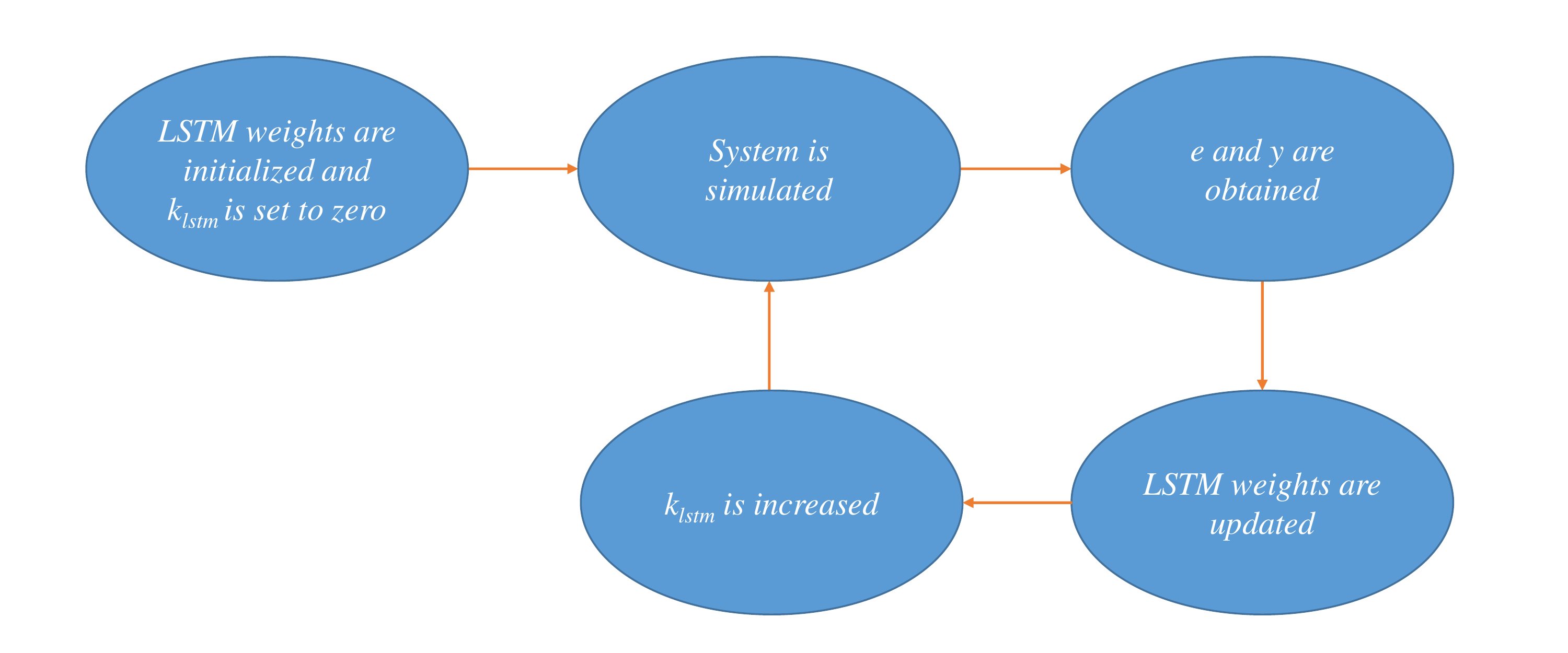}
		\caption{Flow chart of the training process}
		\label{flowchart}
	\end{figure}

	Since there is no initial data set, the data needs to be collected by running the simulation with the initial LSTM weights. Since weight updates of the LSTM network also affect the system dynamics, new data needs to be collected after each episode of training. Before giving the LSTM output $u_{lstm}$ to the system, a gain ``$k_{lstm}$'' scales $u_{lstm}$. The role of this gain is to avoid an untrained LSTM network output from negatively affecting the system at the beginning of the training. $k_{lstm}$ starts from 0 and approaches 1 as the number of training iterations increases, then stays at 1 for the rest of the training. The training loop can be seen in Fig.~\ref{flowchart}.
	
	\subsection{Normalization}
	The inputs of the LSTM network are the components of the tracking error vector $e$, defined in (\ref{tracking_error}). In order to make sure that these components are on a similar scale, normalization needs to be performed. However, since an initial training data set does not exist, data needs to be collected first in order to acquire the parameters for normalization. For collecting the normalization parameters only, the simulation is run with $f_{train}$ (see explanations after (\ref{truth})), $z$ number of times. A gain is uniformly sampled from $[0, 1]$ in every simulation run, to scale $f_{train}$. This random scaling is used only during the normalization parameter collection, not during training.The LSTM network output $u_{lstm}$ is not injected to the system during this process. Within the collected data, the minimum, $e^{min}_i$, and the maximum, $e^{max}_i$, values, where "$i$" is used to refer the $i^{th}$ component, of each component of the error vector, are utilized to obtain the normalized input parameters as
	\begin{equation} \label{input}
		\begin{aligned}
			&e_{norm_i} = (e_i - e^{min}_i)/(e^{max}_i - e^{min}_i).
		\end{aligned}
	\end{equation}

\subsection{Stability Analysis}
The following theorem displays the stability properties of the overall architecture.

\textit{Theorem 1:} Consider the uncertain plant dynamics (\ref{plant dynamics}), subject to Assumptions 1 and 2, and the reference model (\ref{reference model}). Let the control input be defined by (\ref{control input}), which consists of the baseline controller (\ref{baseline}), the ANN controller defined by (\ref{u_ad}), (\ref{fhat}) and (\ref{NN weight updates}), the robustifying term given in (\ref{robustifying_term}) and (\ref{gain}), and the LSTM controller explained in Section \ref{Controller Design}. Then, given that $x_p(0)\in S_p$ (see Assumption 1), the solution $(e(t),\tilde{W}(t),\tilde{V}(t))$ is \textit{uniformly ultimately bounded} (UUB) and converges to a predefined compact set, where $\tilde{V}(t) \triangleq V-\hat{V}(t)$ and $\tilde{W}(t) \triangleq W-\hat{W}(t)$.

\begin{proof} The proof is deferred to the Appendix \ref{theo1}.
	\end{proof}
	\section{Simulations}\label{Simulations}
	In this section, the performance of the proposed control framework is examined using the short-period longitudinal flight dynamics. The short-period dynamics is given as \cite{lavretsky2013robust}
	
	\begin{equation} \label{Short-Period Dynamics}
		\begin{aligned}
			\begin{bmatrix}
				\dot{\alpha} \\ \dot{q} 
			\end{bmatrix} = 
			\begin{bmatrix} \frac{Z_{\alpha}}{mU} \ & 1+ \frac{Z_q}{mU} \\ 	\frac{M_{\alpha}}{I_y} & \frac{M_{q}}{I_y} 
			\end{bmatrix}
			\begin{bmatrix} \alpha \\ q 
			\end{bmatrix}	
			+ \begin{bmatrix} \frac{Z_{\delta}}{mU} \\ \frac{M_{\delta}}{I_y} 
			\end{bmatrix}(u+f(x_p)),
		\end{aligned}
	\end{equation}
	where $\alpha$ is the angle of attack (deg), $q$ is the pitch rate (deg/s), $u$ is the elevator deflection (deg), and $f(x_p)$ is a state dependent matched uncertainty. Elevator magnitude and rate saturation limits are set as $+17/-23$ (deg) and $+37/-37$ (deg/s) \cite{sun2014hybrid}. The commanded input is the pitch rate.
	$Z_{\alpha}$, $Z_q$, $M_{\alpha}$, $M_q$, $Z_{\delta}$ and $M_{\delta}$ are the stability and control derivatives.
	
	The system matrices for a B-747 aircraft flying at the speed of 274 m/s at 6000 m altitude are given as \cite{muthirayan2022memory}
	
	\begin{equation} \label{Ap}
		A_p = \begin{bmatrix} \--0.32 \ & 0.86 \\ -0.93 & -0.43 \end{bmatrix} , B_p = \begin{bmatrix} -0.02 \\ -1.16 \end{bmatrix} , C_p = \begin{bmatrix} 0 \ 1 \end{bmatrix}^T.
	\end{equation}
	The augmented state vector is
	\begin{equation} \label{state}
		x(t) = \begin{bmatrix} x_1(t) & x_2(t) & x_3(t) \end{bmatrix}^T,
	\end{equation}
	where $x_1$ is the angle of attack (rad), $x_2$ is the pitch rate (rad/s) and $x_3$ is the error-integral state (\ref{integral action}).
	The input to the LSTM network is,
	\begin{equation} \label{lstm_input}
		in_{lstm}(t) = \begin{bmatrix} e_{norm_1}(t) & e_{norm_2}(t) & e_{norm_3}(t) \end{bmatrix}^T,
	\end{equation}
	where $e_{norm_1}$, $e_{norm_2}$, and $e_{norm_3}$ are the components of the normalized version of the state tracking error vector (\ref{input}).
	
	The baseline controller (\ref{baseline}) is an LQR controller with cost matrices $Q_{LQR} = I$ and $R_{LQR} = 1$. The nonlinear uncertainty, $f_{train}$ (see explanations after (\ref{truth})) that the Long Short-Term Memory (LSTM) network (\ref{u_lstm}) is trained on is defined as
 \vspace{-1cm}
	\begin{multline}\label{lstm_training_f}
		f_{train}(x)= \\
		\begin{cases}
			0.1x_1x_2 & \text{if } 0\leq t<5\\
			\exp(x_2) & \text{if } 5\leq t<10\\
			2x_1x_2 & \text{if } 10\leq t<15\\
			-0.1\cos(x_1) & \text{if } 15\leq t<20\\
			0.5(x_1x_2) & \text{if } 20\leq t<25\\
			0.1x_1x_2 & \text{if } 25\leq t<30\\
			-x_1x_2(\sin(5x_1x_2)+5\sin({x_2})) & \text{if } 30\leq t<35\\
			x_1x_2(3\sin(2x_1x_2)+2x_1) & \text{if } 35\leq t<45\\
			-x_1x_2(2(\tan(2x_1x_2)+x_2^2) & \text{if } 45\leq t<55\\
			x_1x_2(x_1+x_2) & \text{if } 55\leq t \leq60\\
			\end{cases}.
	\end{multline}
 The LSTM network is trained in the presence of an ANN controller (see Section \ref{ANN}) that has a hidden layer consisting of four neurons. The learning rates in (\ref{NN weight updates_W}) and (\ref{NN weight updates_V}) are set as $F = G = 10$, and the Lyapunov matrix $Q$ in (\ref{lyap eqn}) is set as $Q = I$. The outer weights $(\hat{W})$ and bias $(\hat{b}_w)$ are initialized to zero. The inner weights $(\hat{V})$ and biases $(\hat{b}_v)$ are initialized randomly between 0 and 1. In the simulations, the robustifying gain $k_z$ in (\ref{gain}) and scalar gain $\kappa$ in (\ref{NN weight updates_W}) and (\ref{NN weight updates_V}) are set to 0.

	The LSTM network contains one hidden layer with 128 neurons. The number of neurons in the input layer is 3 due to the number of state-tracking error components given as input to the network (\ref{lstm_input}).
	  Weights are initialized using Xavier Initialization \cite{glorot2010understanding}. The network is trained with stochastic gradient descent, and uses the Adam optimizer with a learning rate of 0.001, an L2 regularization factor of 0.0001, a gradient decay factor of 0.9, and a squared gradient decay factor of 0.999 \cite{kingma2014adam}. The gradient clipping method is used with a threshold value of 1. The simulation step time is set to $0.01$ s. The minibatch size is taken as 1. To cover both low and high-valued uncertainties, $f_{train}$ is scaled by a parameter that took the alternating values of 0.2 and 2.
	
	To obtain the normalization constants given in (\ref{input}), the system is simulated 1000 times ($z = 1000$), with the gain $k_{f}$, (see Section III-B). 
	
	The loss function is chosen as Mean Squared Error (MSE) given as
	\begin{equation}\label{loss}
		L(y,\hat{y}) = \frac{1}{N} \sum_{i = 0}^{N} (y-\hat{y})^2,
	\end{equation}
	where $y$ and $\hat{y}$ are defined in (\ref{lstmtraining}) and (\ref{truth}),and $N$ is the number of data in the data set.
 
	The proposed control framework is tested using the uncertainty defined as 
	\begin{equation}\label{lstm_test_f}
		f_{test}(x)=
		\begin{cases}
			0 & \text{if } 0\leq t<2\\
			-0.1\exp(x_1x_2) & \text{if } 2\leq t<8\\
			0.5x_2^2 & \text{if } 8\leq t<12\\
			0.05\exp({x_1+2x_2}) & \text{if } 12\leq t<20\\
			-0.1\sin(0.05x_1x_2) & \text{if } 20\leq t<28\\
			0.1(x_1^2+x_2^3) & \text{if } 28\leq t<34\\
			-0.1\sqrt{cos(x_2)^2} & \text{if } 34\leq t<40\\
			-0.2\sin(x_1x_2) & \text{if } 40\leq t<49\\
			0.1(x_1x_2)^2 & \text{if } 49\leq t<53\\
			0.5x_2 & \text{if } 53\leq t<60\\
			0.1(|x_2|^{x_1}) & \text{if } 60\leq t<67\\
			-0.5\sin({x_2}) & \text{if } 67\leq t<79\\
			0.01\exp({x_1}) & \text{if } 79\leq t<86\\
			-0.4x_1^2 & \text{if } 86\leq t<98\\
			0.2x_2^2 & \text{if } 98\leq t \leq 110\\
		\end{cases},  
	\end{equation}
	where the function is chosen to have different types of sub-functions with different time intervals, compared to the training uncertainty (\ref{lstm_training_f}).
	
	\subsection{Controller performance in the presence of small uncertainty}	
 
	In this section, the effect of the proposed LSTM augmentation is examined using low-value uncertainties. For this purpose, $f_{test}$ is  scaled by 0.1 during the tests. Then, the tracking performance, tracking error, and control inputs of the closed-loop system are compared with and without LSTM augmentation.

	\begin{figure}[tb]
		\centering
		\includegraphics[width=\linewidth]{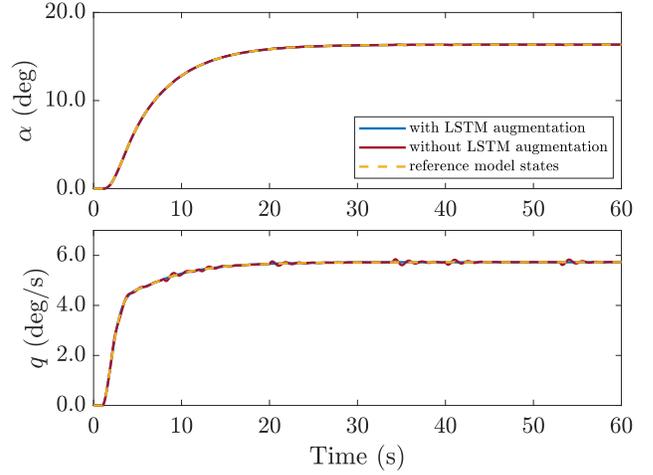}
		\caption{Tracking performances with and without LSTM augmentation, in the presence of small uncertainty.}
		\label{tracking_l}
	\end{figure}

	\begin{figure}[h!]
		\centering
		\includegraphics[width=\linewidth]{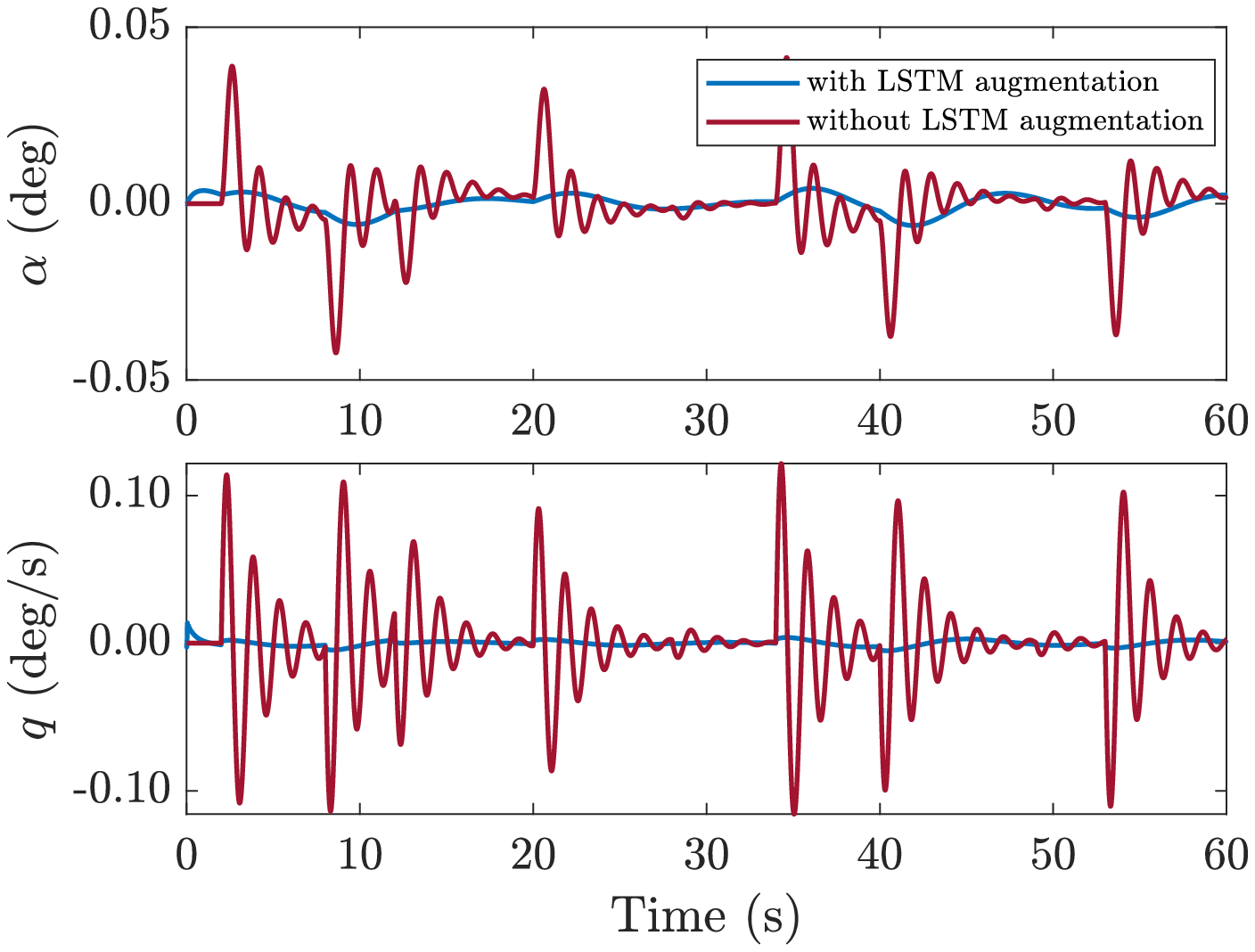}
		\caption{Tracking errors with and without LSTM augmentation, in the presence of small uncertainty.}
		\label{error_l}
	\end{figure}

	\begin{figure}[tb]
		\centering
		\includegraphics[width=\linewidth]{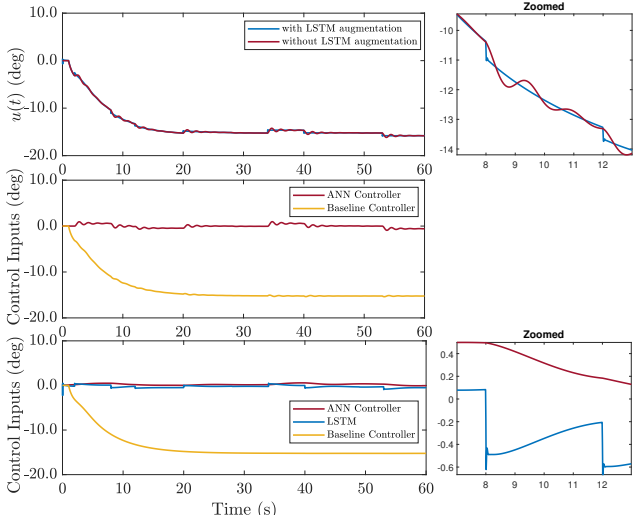}
		\caption{Control inputs in the presence of small uncertainty. Top: Total control input. Middle: The contributions of individual control inputs in the absence of LSTM. Bottom: The contributions of individual control inputs in the presence of LSTM.}
		\label{control_l}
	\end{figure}

    Figures \ref{tracking_l} and \ref{error_l} show the tracking and tracking error curves, respectively. While the error plots demonstrate that LSTM augmentation dramatically reduces both the error magnitudes and the transient oscillations, the error values are small enough to be ignored compared to the absolute values of the states. Therefore, one can conclude that in this scenario LSTM augmented and not-augmented cases show similar performances.   
	In Fig. \ref{control_l}, the control inputs are presented. This figure shows that although LSTM augmentation does not affect the magnitude of the total control input in a meaningful manner, it provides damping to the oscillations by providing small but fast compensation to the introduced uncertainties. As we discussed above, since the ANN controller is already successful in compensating for the uncertainties, in this case, LSTM contribution is not prominent. 
	\subsection{Controller performance in the presence of large uncertainty}
	In this section, the effect of the proposed LSTM augmentation is examined in the presence of high-valued uncertainties, which is obtained by using the $f_{test}$ as is.
	\begin{figure}[h!]
		\centering
		\includegraphics[width=\linewidth]{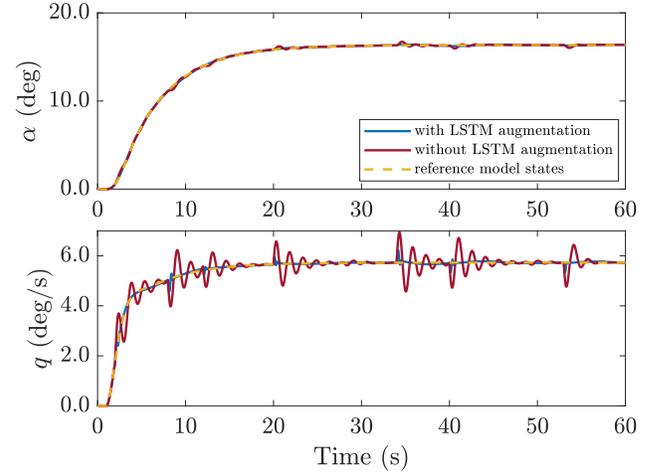}
		\caption{Tracking performances with and without LSTM augmentation, in the presence of large uncertainty.}
		\label{tracking_h}
		\vspace{-4mm}
	\end{figure}
	\hfill
	\begin{figure}[h!]
		\centering
		\includegraphics[width=\linewidth]{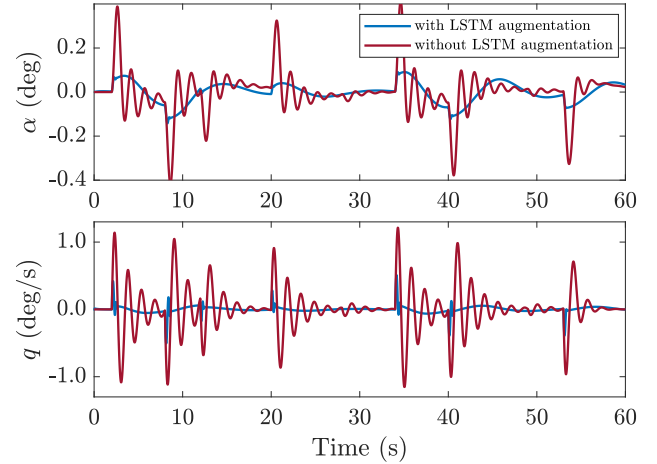}
		\caption{Tracking errors with and without LSTM augmentation, in the presence of large uncertainty.}
		\label{error_h}
		\vspace{-4mm}
	\end{figure}
	
\begin{figure}[h!]
		\centering
		\includegraphics[width=\linewidth]{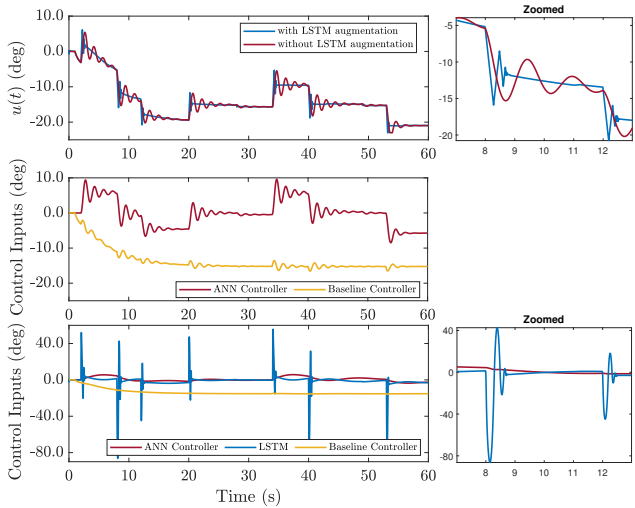}
		\caption{Control inputs in the presence of high uncertainty. Top: Total control input. Middle: The contributions of individual control inputs in the absence of LSTM. Bottom: The contributions of individual control inputs in the presence of LSTM.}
		\label{control_h}
    \end{figure}

	Figures \ref{tracking_h} and \ref{error_h} demonstrate that LSTM augmentation substantially improves the transient response of the system, especially in pitch rate tracking, which is the output of interest (see (\ref{Ap})). It is noted that for this case, the same ANN controller and the LSTM network are used as in the small uncertainty case. Individual control inputs are shown in Fig. \ref{control_h}. In this case, unlike the low-uncertainty case, the LSTM augmentation makes the total control signal observably more agile, which is the main reason why excessive oscillations are prevented. This shows the LSTM network uses its memory to predict high-frequency changes in the system. It is noted that in this large uncertainty case the total control input is saturated. This is evident from Fig. \ref{control_h}, where the middle and the bottom sub-figures show the control signals created by the individual components of the overall controller, while the top sub-figure shows the total control input after saturation. The figure shows that the LSTM network produces large and very fast compensation, which causes saturation. Here, the rate saturation can be problematic since it creates some oscillations in the total control signal (see the zoomed section, blue line, in the top sub-figure). Although the oscillations could be acceptable since they are short-lived, we believe that this issue can also be tackled by training the LSTM network with saturation information. We explain this solution in the next section.

    \subsection{Addressing Saturation}
    
     To handle the saturation issue, we modified LSTM training by a) introducing the saturation limits to the plant dynamics during training, and b) informing the LSTM network whenever the control signal rate-saturates. The latter is achieved by providing additional input to the LSTM network as
    \begin{equation}
	\begin{aligned} \label{rateinf}
		u_{r}(t) = & \begin{cases}
			 0.1 ,  &\text{if } \text{rate saturation is positive}\\
			-0.1,  &\text{if } \text{rate saturation is negative},\\
			0 , &\text{otherwise }
		\end{cases}
	\end{aligned}
    \end{equation}
	
    \begin{figure}[h!]
		\centering
		\includegraphics[width=\linewidth]{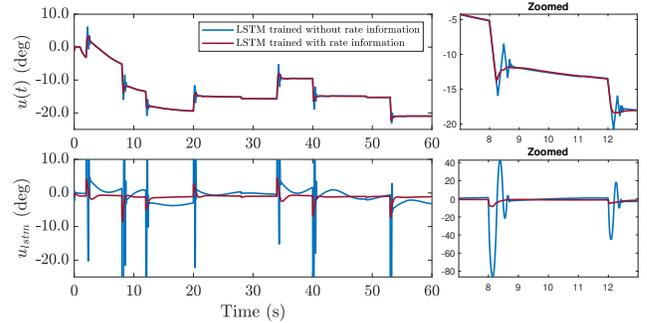}
		\caption{Control signals with an without rate-limit-informed LSTM. Top: Total control input. Bottom: LSTM network contribution to the total control input.}
		\label{rate}
     \end{figure}

    The number of neurons in the input layer of the LSTM network is increased to 4 since the modified input to the LSTM network is
    
    \begin{equation} \label{lstm_input_rate}
		in_{lstm}(t) = \begin{bmatrix} e_{norm_1}(t) & e_{norm_2}(t) & e_{norm_3}(t) & u_r(t) \end{bmatrix}^T.
	\end{equation}

\begin{figure}[h!]
		\centering
		\includegraphics[width=\linewidth]{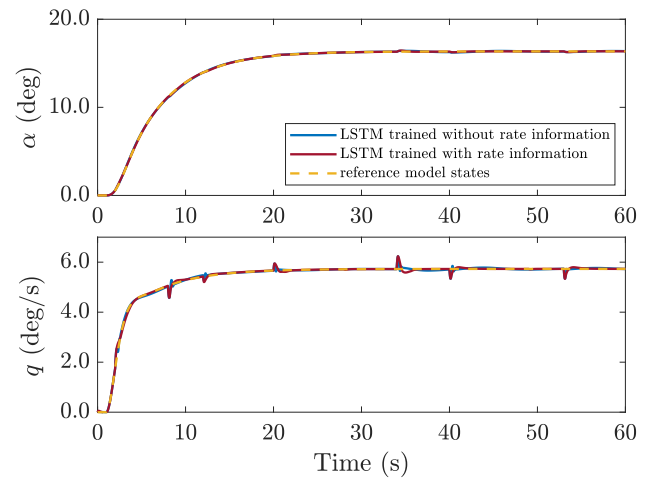}
		\caption{Tracking performances with LSTM trained without rate information and rate-limit-informed LSTM}
		\label{control_h_comp}
    \end{figure}

\begin{figure}[h!]
		\centering
		\includegraphics[width=\linewidth]{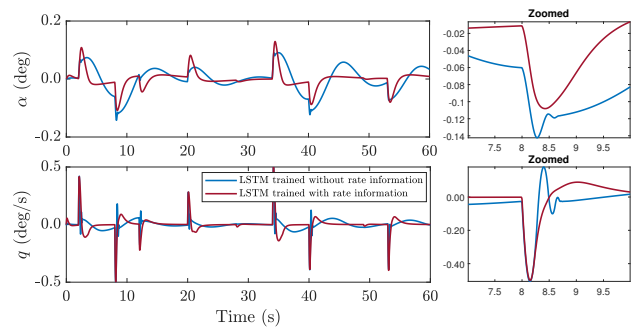}
		\caption{Tracking errors with LSTM trained without rate information and rate-limit-informed LSTM}
		\label{ratecomp}
     \end{figure}

    Figure \ref{rate} shows the LSTM network contribution and the total control inputs, with and without rate-limit information during training.
    It is seen that the LSTM network uses the rate-limit information to provide a smoother compensation. Figures \ref{control_h_comp} and \ref{ratecomp} show the tracking and tracking error curves, respectively, when the LSTM network is trained using the rate-limit information. It is seen that the performance of the proposed controller remains similar for the case when LSTM is not trained using the saturation information, although the pitch rate shows initial small jumps at the instances of uncertainty switches.   

	\section{Summary}\label{Summary}
	In this work, we propose a Long Short-Term Memory (LSTM) augmented adaptive neural network (ANN) control structure to improve the transient response of adaptive closed-loop control systems. We demonstrate that thanks to its time-series prediction capabilities, LSTM helps the ANN controller compensate for the uncertainties in a more agile fashion, resulting in dramatically improved tracking performance.

	\bibliographystyle{IEEEtran}
	\bibliography{IEEEabrv,sample}
	
\appendices \label{Stability}
\section{Error Dynamics and Useful Remarks} \label{dyn}
In this section, the derivation of the ANN controller adaptive laws (\ref{NN weight updates}) and the robustifying term (\ref{robustifying_term}) is shown.

We start by substituting the control input (\ref{control input}) and the baseline controller (\ref{baseline}) into (\ref{augmented_dynamics}), which yields
\begin{equation}\label{p1}
	\begin{aligned}
		\dot{x}(t)=&A_mx(t)+B_mr(t)\\
		&+B(u_{ad}(t)+u_{lstm}(t)+v(t)+f(x_p(t))).
	\end{aligned}
\end{equation}
Using (\ref{tracking_error}), (\ref{p1}) and (\ref{reference model}), the state-tracking error dynamics can be written as
\begin{equation} \label{e1}
	\begin{aligned}
		&\dot{e}(t) =A_me(t)+B(u_{ad}(t)+u_{lstm}(t)+v(t)+f(x_p(t))),
	\end{aligned}
\end{equation}
which, by using the NN approximation property (\ref{approximation}), can be written as
\begin{equation} \label{e2}
	\begin{aligned}
		\dot{e}(t)=&A_me(t)+B(u_{ad}(t)+W^T\sigma(V^Tx_p(t))\\
		&+u_{lstm}(t)+v(t)+\varepsilon{(x_p(t))}).
	\end{aligned}
\end{equation}
The ANN controller $u_{ad}(t)$ is chosen to be the output of a multi-layer NN, as defined in (\ref{u_ad}) and (\ref{fhat}).
Substituting (\ref{u_ad}) and (\ref{fhat}) into (\ref{e2}), yields
\begin{equation} \label{e3}
	\begin{aligned}
		\dot{e}(t)=&A_me(t)+B(-\hat{W}^T(t)\sigma(\hat{V}^T(t)x_p(t))+W^T\sigma(V^Tx_p(t))\\
		&+u_{lstm}(t)+v(t)+\varepsilon{(x_p(t))}),
	\end{aligned}
\end{equation}
where $\hat{V}(t) \in \mathbb{R}^{n_h\times n}$ and $\hat{W}(t) \in \mathbb{R}^{m\times n_h}$ are adaptive NN weights serving as estimates for the unknown ideal weights $V$ and $W$, respectively. The weight estimation errors are defined as
\begin{subequations}\label{tilde}
	\begin{align}
	\tilde{W}(t) =& W-\hat{W}(t), \label{W_tilde} \\
	 \tilde{V}(t) =& V-\hat{V}(t). \label{V_tilde}
	 \end{align}
\end{subequations}
Throughout the remainder of this section, we drop the time dependency notation for simplicity. Further, we denote $\sigma\triangleq\sigma(V^Tx_p(t))$ and $\hat{\sigma}\triangleq \sigma(\hat{V}^Tx_p(t))$. Then, (\ref{e3}) can be written as
\begin{equation} \label{e4}
	\begin{aligned}
		\dot{e}=A_me+B(-\hat{W}^T\hat{\sigma}+W^T\sigma+u_{lstm}+v+\varepsilon{(x_p)}).
	\end{aligned}
\end{equation}
Using (\ref{W_tilde}) in (\ref{e4}) yields
\begin{equation} \label{e5}
	\begin{aligned}
		\dot{e}=A_me+B(\tilde{W}^T\hat{\sigma}+W^T(\sigma-\hat{\sigma})+u_{lstm}+v+\varepsilon{(x_p)}).
	\end{aligned}
\end{equation}

\textit{Remark 4:} Since $r(t)$ is bounded and $A_m$ is Hurwitz, it follows from (\ref{reference model}) that
\begin{equation} \label{bound_on_xr}
	\begin{aligned}
		\norm{x_m(t)} \leq C_m,
	\end{aligned}
\end{equation}
where $C_m>0$ is a known constant, which depends on the user-defined reference input $r(t)$. Therefore, it follows from (\ref{tracking_error}), and the fact that $x(t)\triangleq[x_p(t)^T, x_e(t)^T]^T\in\mathbb{R}^{n}$, that
\begin{equation} \label{bound_on_x}
	\begin{aligned}
		\norm{x_p(t)}\leq\norm{x(t)} \leq C_m+\norm{e(t)},
	\end{aligned}
\end{equation}

\textit{Remark 5:} From the Taylor series expansion
\begin{equation} \label{taylor}
	\begin{aligned}
		\sigma{(V^Tx_p)} = \sigma{(\hat{V}^Tx_p)}+\hat{\sigma}'\tilde{V}^Tx_p+O(\tilde{V}^Tx_p)^2,
	\end{aligned}
\end{equation}
one can write
\begin{equation} \label{t1}
	\begin{aligned}
		\sigma{(V^Tx_p)} - \sigma{(\hat{V}^Tx_p)}=\hat{\sigma}'\tilde{V}^Tx_p+O(\tilde{V}^Tx_p)^2,
	\end{aligned}
\end{equation}
where $\hat{\sigma}'$ is defined in (\ref{sigma}) and $O(\tilde{V}^Tx_p)^2$ denote the higher order terms in the series. Furthermore, since
\begin{equation} \label{t2}
	\begin{aligned}
		O(\tilde{V}^Tx_p)^2=\sigma{(V^Tx_p)} - \sigma{(\hat{V}^Tx_p)}-\hat{\sigma}'\tilde{V}^Tx_p,
	\end{aligned}
\end{equation}
then, for sigmoid and tanh activation functions, $\norm{\sigma(.)}\leq C_\sigma$ and $\norm{\sigma'(.)}\leq C_{\sigma'}$ for known constants $C_\sigma>0$ and $C_{\sigma'}>0$. Hence, using (\ref{bound_on_x}), the higher-order terms in the Taylor series are bounded by
\begin{equation} \label{O bound}
	\begin{aligned}
		\norm{O(\tilde{V}^Tx_p)^2}\leq 2C_\sigma+C_{\sigma'}\norm{\tilde{V}}_FC_m+C_{\sigma'}\norm{\tilde{V}}_F\norm{e}.
	\end{aligned}
\end{equation}

Substituting (\ref{t1}) for $(\sigma-\hat{\sigma})$ in (\ref{e5}), yields
\begin{equation} \label{e6}
	\begin{aligned}
		\dot{e}=&A_me+B(\tilde{W}^T\hat{\sigma}+W^T\hat{\sigma}'\tilde{V}^Tx_p\\
		&+W^TO(\tilde{V}^Tx_p)^2+u_{lstm}+v+\varepsilon{(x_p)}),
	\end{aligned}
\end{equation}
which, by using (\ref{W_tilde}) an (\ref{V_tilde}), can be written as
\begin{equation} \label{Final error dynamics}
	\begin{aligned}
		\dot{e}=&A_me+B(\tilde{W}^T(\hat{\sigma}-\hat{\sigma}'\hat{V}^Tx_p)+\hat{W}^T\hat{\sigma}'\tilde{V}^Tx_p\\
		&w+u_{lstm}+v),
	\end{aligned}
\end{equation}
where
\begin{equation}\label{w}
	w\triangleq \tilde{W}^T\hat{\sigma}'V^Tx_p+W^TO(\tilde{V}^Tx_p)^2+\varepsilon{(x_p)}.
\end{equation}

\textit{Remark 6:} Let 
\begin{equation}\label{Z}
	\begin{aligned}
		Z \triangleq
		\begin{bmatrix}
			W & 0_{n_h\times n_h} \\ 0_{n_p\times m} & V
		\end{bmatrix}\in \mathbb{R}^{(n_h+n_p)\times (m+n_h)}.
	\end{aligned}
\end{equation}
Using (\ref{Z}) and assumption 2, one can write
\begin{equation} \label{Znorm}
	\norm{Z}_F^2 = \norm{W}_F^2+\norm{V}_F^2\leq W_M^2+V_M^2,
\end{equation}
and hence, $\norm{Z}_F\leq Z_M$, where $Z_M\triangleq \sqrt{W_M^2+V_M^2}$. Furthermore, it follows from (\ref{Znorm}) that $\norm{W}_F^2 \leq Z_M$ and $\norm{V}_F^2 \leq Z_M$. Similarly, by defining 
\begin{equation}\label{Zhat}
	\begin{aligned}
		\hat{Z}(t) \triangleq
		\begin{bmatrix}
			\hat{W}(t) & 0_{n_h\times n_h} \\ 0_{n_p\times m} & \hat{V}(t)
		\end{bmatrix}\in \mathbb{R}^{(n_h+n_p)\times (m+n_h)}, 
	\end{aligned}
\end{equation}
one can show that $\norm{\tilde{W}(t)}_F\leq\norm{\tilde{Z}(t)}_F$ and $\norm{\tilde{V}(t)}_F\leq\norm{\tilde{Z}(t)}_F$, where $\tilde{Z}(t)\triangleq Z-\hat{Z}(t)$.

\textit{Remark 7:} Using (\ref{eps}), (\ref{O bound}) and Remark 6, one can write
\begin{equation} \label{w bound}
	\begin{aligned}
	\norm{w}\leq& \norm{\tilde{Z}}_FC_{\sigma'}Z_M(C_m+\norm{e})\\
	&+Z_M( 2C_\sigma+C_{\sigma'}\norm{\tilde{Z}}_FC_m+C_{\sigma'}\norm{\tilde{Z}}_F\norm{e})+\varepsilon_N,
	\end{aligned}
\end{equation}
or
\begin{equation} \label{w bound}
	\begin{aligned}
		\norm{w}\leq& C_0+C_1\norm{\tilde{Z}}_F+C_2\norm{\tilde{Z}}_F\norm{e},
	\end{aligned}
\end{equation}
where
\begin{subequations}
	\begin{align}
		C_0\triangleq&2C_\sigma Z_M+\varepsilon_N,\\
		C_1\triangleq& 2C_{\sigma'}C_mZ_M,\\
		C_2\triangleq&2C_{\sigma'}Z_M.
	\end{align}
\end{subequations}

	\textit{Remark 8:} Sigmoid and tanh activation functions help define a bound on the LSTM output: We can write (\ref{hidden state}) as
	\begin{equation}
		h^n = 
		\begin{pmatrix}
			{\sigma_g}_1(\dotsm){\sigma_c}_1(\dotsm) \\
			\vdots \\
			{\sigma_g}_{n_h}(\dotsm){\sigma_g}_{n_h}(\dotsm)
		\end{pmatrix} \in \mathbb{R}^{n_h},
	\end{equation}
	where $n_h$ is the number of neurons in the last hidden layer.
	Then,
	\begin{equation} \label{h_norm}
		\begin{aligned}
			&\norm{h^n}^2 = \sum_{i}^{n_h}{\sigma^2_g}_i(\dotsm){\sigma^2_c}_i(\dotsm) \leq n_h.
		\end{aligned}
	\end{equation}
	Hence,
	\begin{equation} \label{bound_h}
		\norm{h^n} \leq \sqrt{n_h}.
	\end{equation}
	The bound on the LSTM output can be defined as
	\begin{equation} \label{bound_lstm}
		\norm{u^n_{lstm}} \leq \norm{W_{fc}}_F\sqrt{n_h}+\norm{b_{fc}} \triangleq \bar{u}_{lstm},
	\end{equation}
which by using (\ref{D2A}) implies that $\norm{u_{lstm}(t)}\leq\bar{u}_{lstm}$.

\section{Proof of Theorem 1:} \label{theo1}
Let the approximation property (\ref{approximation}) hold on a known compact set $S_p\triangleq \{x_p:\norm{x_p}\leq b_x\}$ as stated in Assumption 1, for some $b_x>C_m$. Defining
\begin{equation}\label{Sp}
	\begin{aligned}
		S \triangleq \{x:\norm{x}\leq b_x\}, \quad b_x>C_m,
	\end{aligned}
\end{equation}
and using the fact that $\norm{x_p}\leq\norm{x}$, $x\in S$ implies that $x_p\in S_p$. Consider the compact set
\begin{equation}\label{lyapunov}
	\begin{aligned}
		S_e \triangleq \{e:\norm{e}\leq b_x-C_m\},
	\end{aligned}
\end{equation}
which imply that once $e\in S_e$, then $x_p\in S_p$. Let $e(0) \in S_e$. Then, $x_p(0)\in S_p$ and (\ref{approximation}) holds. The proof proceeds by showing that $e(t)\in S_e$, $\forall t\geq0$.

Consider the Lyapunov function
\begin{equation}\label{V}
	\begin{aligned}
		V = e^TPe+\text{tr}\{\tilde{W}^TF^{-1}\tilde{W}\}+\text{tr}\{\tilde{V}^TG^{-1}\tilde{V}\}.
	\end{aligned}
\end{equation}
Differentiating (\ref{V}) along the trajectory (\ref{Final error dynamics}) yields,
\begin{equation}\label{v1}
	\begin{aligned}
		\dot{V} =& 
		\dot{e}^TPe+e^TP\dot{e}+2\text{tr}\{\tilde{W}^TF^{-1}\dot{\tilde{W}}\}+2\text{tr}\{\tilde{V}^TG^{-1}\dot{\tilde{V}}\}\\
		=&e^TA_m^TPe+e^TPA_me+2e^TPB\tilde{W}^T(\hat{\sigma}-\hat{\sigma}'\hat{V}^Tx_p)\\
		&+2e^TPB\hat{W}^T\hat{\sigma}'\tilde{V}^Tx_p+2e^TPB(w+u_{lstm}+v)\\
		&+2\text{tr}\{\tilde{W}^TF^{-1}\dot{\tilde{W}}\}+2\text{tr}\{\tilde{V}^TG^{-1}\dot{\tilde{V}}\}.\\
	\end{aligned}
\end{equation}
Using (\ref{lyap eqn}) and $\text{tr}\{A_1A_2\}=\text{tr}\{A_2A_1\}$ yields
\begin{equation}\label{v2}
	\begin{aligned}	
		\dot{V}=&-e^TQe+2\text{tr}\{\tilde{W}^T(\hat{\sigma}-\hat{\sigma}'\hat{V}^Tx_p)e^TPB\}\\
		&+2\text{tr}\{\tilde{V}^Tx_pe^TPB\hat{W}^T\hat{\sigma}'\}+2e^TPB(w+u_{lstm}+v)\\
		&+2\text{tr}\{\tilde{W}^TF^{-1}\dot{\tilde{W}}\}+2\text{tr}\{\tilde{V}^TG^{-1}\dot{\tilde{V}}\}.\\
	\end{aligned}
\end{equation}
It follows from (\ref{tilde}) that $\dot{\tilde{W}}=-\dot{\hat{W}}$. Substituting (\ref{NN weight updates}) into (\ref{v2}) yields
\begin{equation}\label{v3}
	\begin{aligned}
		\dot{V} = 
		&-e^TQe+2e^TPB(w+u_{lstm}+v)\\
		&+2\text{tr}\{\tilde{W}^T\kappa\norm{e}\hat{W}\}+2\text{tr}\{\tilde{V}^T\kappa\norm{e}\hat{V}\}.
	\end{aligned}
\end{equation}
Using (\ref{tilde}), (\ref{Z}), and (\ref{Zhat}), one can write 
\begin{equation}\label{v4}
	\begin{aligned}
		\dot{V} = 
		&-e^TQe+2e^TPB(w+u_{lstm}+v)\\
		&+2\kappa\norm{e}\text{tr}\{\tilde{W}^T(W-\tilde{W})\}+2\kappa\norm{e}\text{tr}\{\tilde{V}^T(V-\tilde{V})\}\\
		=&-e^TQe+2e^TPB(w+u_{lstm}+v)\\
		&+2\kappa\norm{e}\text{tr}\{\tilde{Z}^T(Z-\tilde{Z})\}.
	\end{aligned}
\end{equation}
Since
\begin{equation}
	\begin{aligned}
	\text{tr}\{\tilde{Z}^T(Z-\tilde{Z})\}=& \text{tr}\{\tilde{Z}^TZ\}-\norm{\tilde{Z}}_F^2\\
	&\leq\norm{\tilde{Z}}_F(Z_M-\norm{\tilde{Z}}_F),
	\end{aligned}
\end{equation}
it follows from (\ref{v4}) that
\begin{equation}\label{v5}
	\begin{aligned}
		\dot{V} \leq 
		&-\lambda_{min}(Q)\norm{e}^2+2\norm{B^TPe}(\norm{w}+\norm{u_{lstm}})\\
		&+2e^TPBv+2\kappa\norm{e}\norm{\tilde{Z}}_F(Z_M-\norm{\tilde{Z}}_F),
	\end{aligned}
\end{equation}
which, by using (\ref{w bound}) and Remark 8, can be written as
\begin{equation}\label{v6}
	\begin{aligned}
		\dot{V} \leq 
		&-\lambda_{min}(Q)\norm{e}^2+2\norm{B^TPe}(C_0+\bar{u}_{lstm})\\
		&+2C_1\norm{B^TPe}\norm{\tilde{Z}}_F+2C_2\norm{B^TPe}\norm{e}\norm{\tilde{Z}}_F\\
		&+2e^TPBv+2\kappa\norm{e}\norm{\tilde{Z}}_F(Z_M-\norm{\tilde{Z}}_F).
	\end{aligned}
\end{equation}
The proposed robustifying term (\ref{robustifying_term}) is designed to cancel the $4^\text{th}$ term of (\ref{v6}). Two cases follow from using (\ref{robustifying_term}) in (\ref{v6}): \textit{a)} $\norm{B^TPe}=0$ and \textit{b)} $\norm{B^TPe}\neq0$.

\textit{Case a)} $\norm{B^TPe}=0$: In this case, $v=0$, and (\ref{v6}) reduces to
\begin{equation}\label{v7a}
	\begin{aligned}
		\dot{V} \leq 
		&-\lambda_{min}(Q)\norm{e}^2+2\kappa\norm{e}\norm{\tilde{Z}}_F(Z_M-\norm{\tilde{Z}}_F)\\
		=&-\norm{e}(\lambda_{min}(Q)\norm{e}+2\kappa\norm{\tilde{Z}}_F(\norm{\tilde{Z}}_F-Z_M))\\
		=&-\norm{e}\Big(\lambda_{min}(Q)\norm{e}+2\kappa(\norm{\tilde{Z}}_F-\frac{Z_M}{2})^2-\kappa\frac{Z_M^2}{2}\Big),
	\end{aligned}
\end{equation}
which implies that $\dot{V}<0$ if either
\begin{equation}\label{a) e bound}
	\norm{e}>\frac{\kappa Z_M^2}{2\lambda_{min}(Q)},
\end{equation}
or
\begin{equation}\label{a) Z bound}
	\norm{\tilde{Z}}_F>Z_M.
\end{equation}

\textit{Case b)} $\norm{B^TPe}\neq0$: Then, by using $\norm{\tilde{Z}}_F\leq\norm{\hat{Z}}_F+Z_M$, and substituting (\ref{robustifying_term}) into (\ref{v6}), it follows that
\begin{equation}\label{v7b}
	\begin{aligned}
		\dot{V} \leq 
		&-\lambda_{min}(Q)\norm{e}^2+2\norm{B^TPe}(C_0+\bar{u}_{lstm})\\
		&+2C_1\norm{B^TPe}\norm{\tilde{Z}}_F\\
		&+2C_2\norm{B^TPe}\norm{e}(\norm{\hat{Z}}_F+Z_M)\\
		&+2e^TPBv+2\kappa\norm{e}\norm{\tilde{Z}}_F(Z_M-\norm{\tilde{Z}}_F)\\
		=&-\lambda_{min}(Q)\norm{e}^2+2\norm{B^TPe}(C_0+\bar{u}_{lstm})\\
		&+2C_1\norm{B^TPe}\norm{\tilde{Z}}_F+2\kappa\norm{e}\norm{\tilde{Z}}_F(Z_M-\norm{\tilde{Z}}_F)\\
		&+2C_2\norm{B^TPe}\norm{e}(\norm{\hat{Z}}_F+Z_M)\\
		&-2k_z\norm{B^TPe}\norm{e}(\norm{\hat{Z}}_F+Z_M).
	\end{aligned}
\end{equation}
Since $k_z>C_2$, (\ref{v7b}) reduces to 
\begin{equation}\label{v8b}
	\begin{aligned}
		\dot{V} \leq 
		&-\lambda_{min}(Q)\norm{e}^2+2\norm{B^TPe}(C_0+\bar{u}_{lstm})\\
		&+2C_1\norm{B^TPe}\norm{\tilde{Z}}_F+2\kappa\norm{e}\norm{\tilde{Z}}_F(Z_M-\norm{\tilde{Z}}_F)\\
		\leq& -\lambda_{min}(Q)\norm{e}^2+2\norm{B^TP}_F\norm{e}(C_0+\bar{u}_{lstm})\\
		&+2C_1\norm{B^TP}_F\norm{e}\norm{\tilde{Z}}_F+2\kappa\norm{e}\norm{\tilde{Z}}_F(Z_M-\norm{\tilde{Z}}_F)\\
		=&-\norm{e}\Big(\lambda_{min}(Q)\norm{e}-2\norm{B^TP}_F(C_0+\bar{u}_{lstm})\\
		&-2C_1\norm{B^TP}_F\norm{\tilde{Z}}_F-2\kappa\norm{\tilde{Z}}_F(Z_M-\norm{\tilde{Z}}_F)\Big)\\
		=&-\norm{e}\Big(\lambda_{min}(Q)\norm{e}+2\kappa(\norm{\tilde{Z}}_F-\frac{q_1}{2})^2-\kappa\frac{q_1^2}{2}-2q_0\Big),
	\end{aligned}
\end{equation}
where $q_0\triangleq \norm{B^TP}_F(C_0+\bar{u}_{lstm})$ and $q_1\triangleq C_1\norm{B^TP}_F/\kappa+Z_M$. Therefore, it follows that $\dot{V}<0$ if either
\begin{equation}\label{b) e bound}
	\norm{e}>\frac{\kappa q_1^2+4q_0}{2\lambda_{min}(Q)},
\end{equation}
or
\begin{equation}\label{b) Z bound}
	\norm{\tilde{Z}}_F>\frac{q_1}{2}+\sqrt{\frac{q_1^2}{4}+\frac{q_0}{\kappa}}.
\end{equation}

Hence, using the fact that the bounds given in (\ref{b) e bound}) and (\ref{b) Z bound}) are bigger than the bounds in (\ref{a) e bound}) and (\ref{a) Z bound}), and by selecting the Lyapunov matrix $Q$ such that
\begin{equation}
	\lambda_{min}(Q)>\frac{\kappa q_1^2+4q_0}{2(b_x+C_m)},
\end{equation} 
it can be shown that the solution $(e(t),\tilde{Z}(t))$ is \textit{uniformly ultimately bounded} (UUB) \cite{Khalil}, and converges to the compact set
\begin{equation}\label{E}
	\begin{aligned}
	E\triangleq \Big\{(e,\tilde{Z}):& \norm{e}\leq\frac{\kappa q_1^2+4q_0}{2\lambda_{min}(Q)}<b_x+C_m\\
	& \text{and}\:\: \norm{\tilde{Z}}_F\leq\frac{q_1}{2}+\sqrt{\frac{q_1^2}{4}+\frac{q_0}{\kappa}}\Big\}.
	\end{aligned}
\end{equation}
This implies that, since $e(0)\in S_e$ and $\dot{V}<0$ on the boundary of $S_e$, $e(t)$ stays within $S_e$ for all $t\geq0$. Therefore $x_p(t)\in S_p$ for all $t\geq0$.

Since the reference input $r(t)$ is bounded, $x_m(t)$ is bounded. Therefore, the boundedness of $e(t)=x(t)-x_m(t)$ implies that $x(t)$, and therefore all system signals, are bounded. \hfill $\blacksquare$
\end{document}